# Planning for Rhythmized Urban Parks: Temporal Park Classification and Modes of Action

Xiyuan Ren [a,b], ChengHe Guan [a,c] *, Shengze Chen [a], Meizi You [a,d], Ying Li [a,c], Kangning Huang [a,c]

[a] *Shanghai Key Laboratory of Urban Design and Urban Science, NYU Shanghai, Shanghai, China*

[b] *Tandon School of Engineering, New York University, New York, USA*

[c] *Division of Arts and Sciences, NYU Shanghai, Shanghai, China*

[d] *School of Ecological and Environmental Sciences, East China Normal University, Shanghai, China*

* corresponding author: ChengHe Guan cg157@nyu.edu

**Problem, research strategy, and findings:** Urban parks offer residents significant physiological and mental health benefits, improving their quality of life. However, traditional park planning often treats parks as static spatial resources, neglecting their temporal dimension of visitation rhythms. This study proposes a paradigm shift in classifying, programming, and designing parks. Utilizing 1.5 million mobile phone records, we classified 254 urban parks in Tokyo based on their visitation patterns across different times of the day, week, and year. Our results showed that parks are rhythmized by seasonal events and daily activities, exhibiting complex visiting patterns shaped by the combined effects of preference variation and accessibility barriers. We concluded by discussing modes of action for temporal park planning practice.

**Takeaway for practice:** Park planners, designers, and policymakers should seek to incorporate temporality in activity-based park design and programming. This can be fulfilled in four ways by (1) tracing year-round, citywide park activities through data-driven methods, (2) implementing temporary and tactical designs for dynamic park demands, (3) establishing an inclusive park system that helps improve spatiotemporal equity, and (4) encouraging public engagement that cultivates the sense of time and identity.

With accelerating urbanization and the speeding pace of daily life, cities face heightened diversity and temporality in human activities (Lynch, 1976; Wunderlich, 2013). In response, the concepts of rhythmanalysis (Lefebvre, 2004), urban chronopolis (Osman & Mulíček, 2017), and place-temporality (Wunderlich, 2023) have gained wide attention, reimagining urban spaces as dynamic systems synchronized with the temporal rhythms of urban life. Urban parks, as critical sites for physical activity and social interaction, are increasingly rhythmized by the ebb and flow of park visits (Veitch et al., 2021; Wang et al., 2021). Such temporal complexity challenges traditional place-based park planning, which prioritizes static attributes like size and location (Ibes, 2015). The mismatch between rigid infrastructure and fluid usage patterns underscores the urgency of rethinking how parks are programmed and designed in rhythm-driven cities.

The advent of geolocation techniques has transformed urban park studies, enabling deeper analyses of park usage, visitor behaviors, and temporal trends through real-time, granular data (Arti & Jumadi, 2024; Guan et al., 2020; Song et al., 2024). While studies increasingly recognize the importance of temporality in park planning, critical gaps persist in the academic and practical discourse on rhythmized parks. First, classifications of urban parks remain overwhelmingly place-based, focusing on static attributes like size, location, and fixed amenities while neglecting temporal dynamics such as daily usage peaks and seasonal variability (Zhang et al., 2021; Zhou et al., 2024). Second, practices of temporal park design are often siloed, limited to isolated case studies in specific neighborhoods or districts (Kotsila et al., 2020; Petryshyn, 2022), without systemic frameworks to harmonize temporal demands across a city. The lack of innovative park classifications and scalable planning strategies undermines efforts to address park temporality at a systemic level.

This paper seeks to address these gaps through a dual focus: temporal park classification and modes of actions. First, we summarize the theoretical and practical underpinnings of temporal park planning, synthesizing insights from time-geography, temporal urbanism, and rhythmized urban parks and calling for a paradigm shift towards temporal park classification. Second, using a case study in Tokyo, we propose a novel classification framework for urban parks using mobile phone data. The framework integrates daily, weekly, and seasonal variations in park visitation patterns to capture temporal dynamics alongside spatial attributes. Third, we demonstrate how this classification model uncovers behavioral insights—such as visitation frequency, visitor composition, travel distance, and stay duration—that traditional approaches fail to capture. Finally, we discuss modes of action for temporal park planning practice from four aspects: (1) data-driven tracking of park visitation rhythms, (2) temporary design for dynamic needs, (3) spatiotemporal equity in park systems, and (4) community-driven sense of time.

The remainder of this paper is structured as follows. "The Need for a Paradigm Shift" section introduces the theoretical and practical backgrounds of temporal urban park planning. "Temporal Park Classification Using Mobile Phone Data in Tokyo" section proposes our data-driven methodology, leveraging anonymized mobile phone data to capture daily, weekly, and seasonal park rhythms. "Park Classification Results and Behavioral Insights" section presents empirical findings in Tokyo, revealing temporal usage patterns that challenge conventional park typologies. Finally, "Modes of Action for Temporal Park Planning" section emphasizes data-driven techniques, temporary design, equitable programming, and community co-creation. The conclusion synthesizes these insights, outlining pathways for future research and practice in temporal park planning.

**The Need for A Paradigm Shift**

***Urban Temporality: Theoretical and Practical Underpinnings***

Our cities are recognized not merely as spatial artifacts but as dynamic temporal constructs, where time profoundly influences human experiences, equity, and sustainability (Lefebvre, 2013; Osman & Mulíček, 2017; Wunderlich, 2023). This temporal turn challenges traditional planning paradigms prioritizing static spatial solutions over adaptive, time-sensitive strategies. At its core lies a theoretical reorientation toward the temporality, rhythms, and cycles of urban life (Liu et al., 2021; Mulíček et al., 2015).

*Time-geography* proposed by Hägerstrand (1970) maps how individuals navigate spatial-temporal constraints, such as fixed work hours or mobility barriers, through the framework of "time-space prisms". This framework highlights the inequities embedded in rigid urban systems, where marginalized groups—such as shift workers or caregivers—often face temporal exclusion. Complementing this, Lefebvre's concept of ***rhythmanalysis*** (2004) provides a time lens, framing cities as palimpsests of overlapping rhythms—natural (daylight, seasons), social (work schedules, cultural rituals), and mechanical (traffic flows, infrastructure operations). Further expanding this course, Osman and Mulíček (2017) introduced the notion of ***urban chronopolis***, reimagining cities as assemblages of dislocated, rhythmized places. Wunderlich (2013) combined spatial and temporal aspects by examining ***place-temporality***, i.e., how the synchronization of pedestrian flows with public transport schedules shapes urban vitality. She further emphasized the need for temporal urban design, advocating for spaces that accommodate both fast-paced urban rhythms and slower, contemplative interactions (Wunderlich, 2023).

These theoretical advances have catalyzed innovative planning practices that prioritize temporality. ***Temporary urbanism***, for instance, employs short-term interventions—such as pop-up parks, seasonal markets, or interim land uses—to stimulate creativity and adapt to evolving urban needs (Stevens et al., 2024). Similarly, ***tactical urbanism*** leverages grassroots, low-cost projects—such as guerrilla gardening or parklet installations—to democratize urban experimentation, enabling communities to reshape spaces in real time (Bråten, 2024). At the policy level, cities like Barcelona and Vienna have implemented the concept of ***15-Minute City (FMC)*** or ***chrono-urbanism***, where temporal zoning regulations are integrated to balance day-night rhythms, mitigate noise conflicts, and ensure equitable access to public spaces after dark (Ferrer-Ortiz et al., 2022). The integration of temporal theories and practices offers transformative insights for planning cities and their sub-systems.

### *Rhythmized Urban Parks: Practices and Challenges*

Urban parks are not static landscapes but dynamic spaces rhythmized by the ebb and flow of human activity (Ren & Guan, 2022; Wolch et al., 2014). Unlike the rigid, clock-driven schedules of work and daily routines, urban parks offer a unique opportunity to reconnect with natural and social rhythms, fostering a deeper awareness of time (Jabbar et al., 2022). Daily, weekly, and seasonal patterns—such as morning joggers, weekend family picnics, or summer festivals—imprint distinct temporal signatures on park use, reflecting the pace of urban life (Veitch et al., 2021; Wang et al., 2021). Such temporal variability underscores that parks should be living systems, continuously adapting to the temporalities of their users.

There is a growing emphasis on integrating temporal elements into park design to create dynamic and adaptable public spaces. For instance, in New York City, various neighborhoods have implemented temporary festivals, markets, art installations, and lighting demonstrations in parks, lasting from a few months to nearly a year (Erdmann-Goldoni, 2024). Temporary gardens in cities like Liverpool, Barcelona, and Dublin provide temporal layers to urban landscapes where public artworks are displayed in open-air settings (Kotsila et al., 2020; Martin et al., 2019). In North California, winter-friendly designs, such as heated benches, sheltered walkways, and seasonal lighting, have been implemented to increase year-round public space usage, promoting pedestrian interaction and environmental awareness (Petryshyn, 2022). These practices underscore a shift towards incorporating temporality in local urban park design.

However, significant challenges stand in the way of planning rhythmized parks as a cohesive system across the entire city. First, accommodating diverse demands across time and user groups requires reconciling conflicting needs—such as quiet zones for relaxation on weekdays versus event spaces for gatherings on holidays (Zhai et al., 2021). This complexity is compounded by socio-cultural differences in park usage, such as intergenerational divides in activity timing or disparities in access to leisure time (Loukaitou-Sideris et al., 2016). Second, mitigating peaks of overcrowding and valleys of underuse requires creative strategies to redistribute park activities. For example, parks adjacent to office districts may experience midday overcrowding during lunch hours but become nearly deserted in the evenings and weekend (Chen & Hedayati Marzbali, 2024). Third, ensuring spatial-temporal equity remains a critical hurdle. The uneven distribution of temporal resources—such as access to leisure time, mobility services, and operating facilities—continues to constrain the recreational opportunities of marginal communities (Rigolon & Németh, 2021). A critical step in addressing these challenges is to establish an innovative park classification framework based on residents' spatiotemporal park activities.

### *A Critical Step: Temporal Park Classification*

Park classification serves as a fundamental tool in park planning and management, providing a structured framework for understanding the diversity of parks and their roles within the urban environment (Loukaitou-Sideris et al., 2016; Pfeiffer et al., 2020; Zhou et al., 2021). Traditional park classification has long centered on place-based approaches, which prioritize physical attributes such as size, location, and designated functions (Brown et al., 2014; Chen et al., 2018; Chuang et al., 2022). While scholars have proposed more nuanced frameworks—integrating physical, spatial, and built-environment criteria to address equity gaps (Ibes, 2015)—these remain rooted in static, place-centric paradigms.

Planning with place-based park categories suffers from critical limitations. By fixating on physical form, it often neglects the social and temporal contexts that define how parks are experienced (Zhang et al., 2021; Zhou et al., 2024). Moreover, place-based approaches fail to account for how park functions shift across time: a daytime jogging route may transform into an evening social hub, serving distinct user groups with divergent needs  (Guan et al., 2021; Zhai et al., 2021). These gaps underscore the inadequacy of place-based models in addressing the dynamic, multi-functional realities of urban parks, necessitating a shift toward activity-based approaches that prioritize how and when parks are used over their static physical traits.

Advances in location-aware technologies over the past decade have enabled unprecedented access to spatial-temporal data, igniting interest in activity-based park studies (Guan & Zhou, 2024; Liang & Zhang, 2021). For instance, Monz et al. (2019) examined year-round GPS data from 22 parks in California, suggesting that peak arrivals occurred between 9 a.m. and 10 a.m. on the weekends, while weekday use was lower and temporally different. Walter et al. (2023) employed social media data to analyze park use in Philadelphia across various demographic groups and illustrated how behaviors in social media can uncover disparities in access and use that traditional categories might overlook. Zhou et al. (2024) utilized over 5.9 million mobile phone records in Tokyo and classified urban parks based on four indicators: activity intensity, utilization efficiency, temporal occupancy, and revisit volume.

However, a significant research gap remains in integrating daily, weekly, and seasonal intricacies in park visitation. To date, few studies have incorporated citywide, year-round visitation dynamics into park classification, and even fewer have proposed strategies adaptive to temporal park categories. The following sections fill the gap by introducing an innovative park classification approach using mobile phone data. Based on a case study in Tokyo, we illustrate how this classification model reveals behavioral insights that traditional ones may overlooked and discuss how to incorporate these insights in future park planning and design.

**Temporal Park Classification Using Mobile Phone Data in Tokyo**

Figure 1 presents the workflow of temporal park classification using mobile phone data in Tokyo, which encompasses two steps: (1) park selection and data collection, and (2) analytical methods for temporal classification.

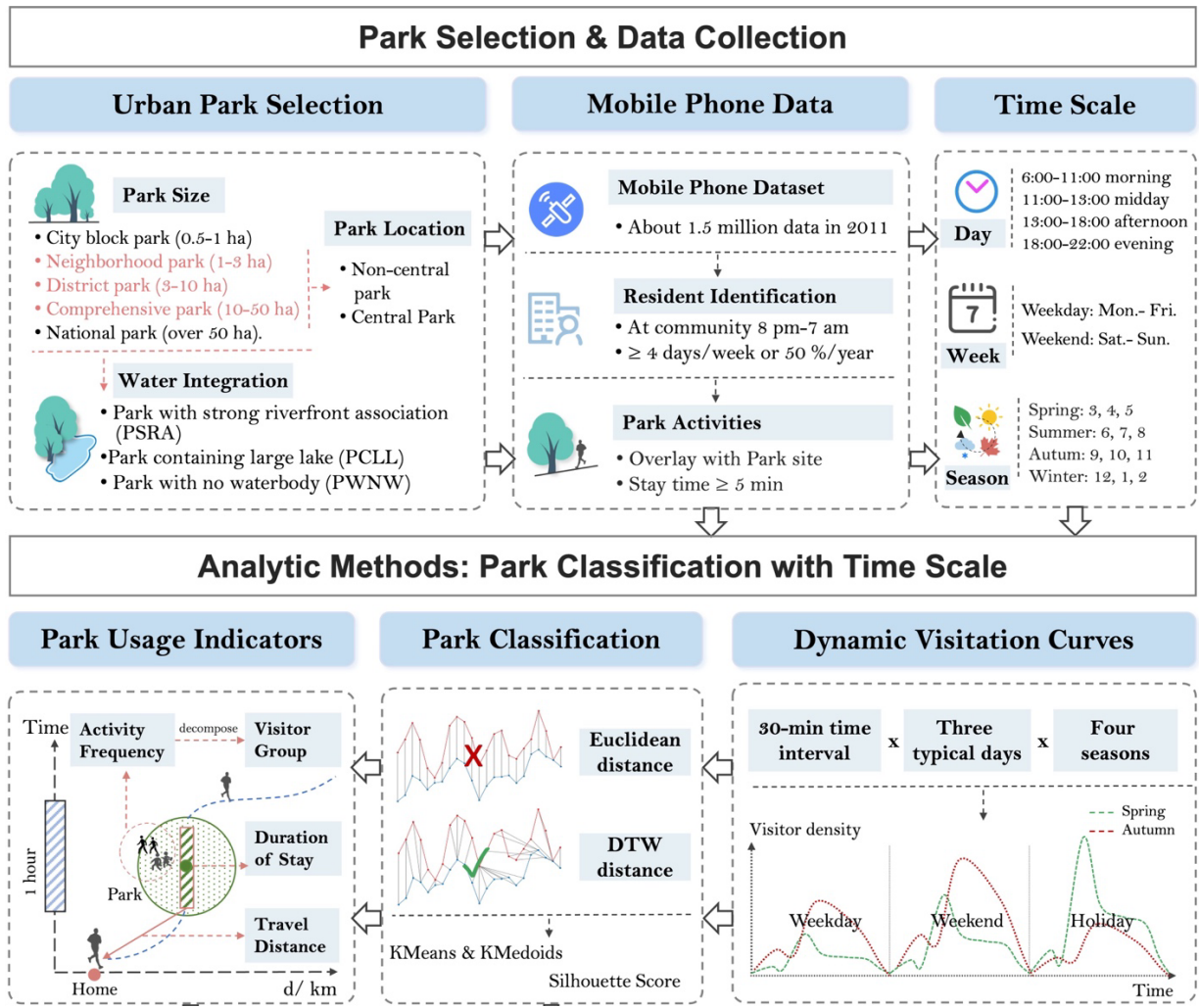

Figure 1. Temporal park classification framework.

## *Park Selection and Data Collection*

### *Study area*

We focus on Tokyo's 23 special-ward area, which is the political, economic, and cultural hub of Japan. These wards cover an total area of 628km², housing 9.78 million residents—a population density of about 15.58 thousand persons per km²—as of 2023 (Tokyo Metropolitan Government, 2023). The high population density and fast-paced lifestyle make this area an ideal case for examining temporality in park activities.

Japan's Urban Park Law classifies parks into five categories and suggests recommended sizes for each: city block park (0.5-1ha), neighborhood park (1-3ha), district park (3-10ha), comprehensive park (10-50ha), and national park (over 50ha) (Setagaya Ward Government, 2016). We focus on neighborhood, district, and comprehensive parks since they are functionally diverse and typically designed for daily activities. Figure 2(a) shows the geographic distribution of the 254 selected parks. We define the central area as the five core wards (Chiyoda, Chuo, Minato, Shinjuku, Shibuya) along with a 1500m buffer zone around Yamanote Line (Ke et al., 2021), as shown in Figure 2(b). Following Grilli et al. (2020)'s work, parks are categorized into three types regarding waterbody integration (Figure 2(c)): (1) parks without waterbody ; (2) parks containing large lakes; and (3) parks with riverfront association. Detailed park counts by size, location, and water integrations are provided in Table S1 in Appendix A.

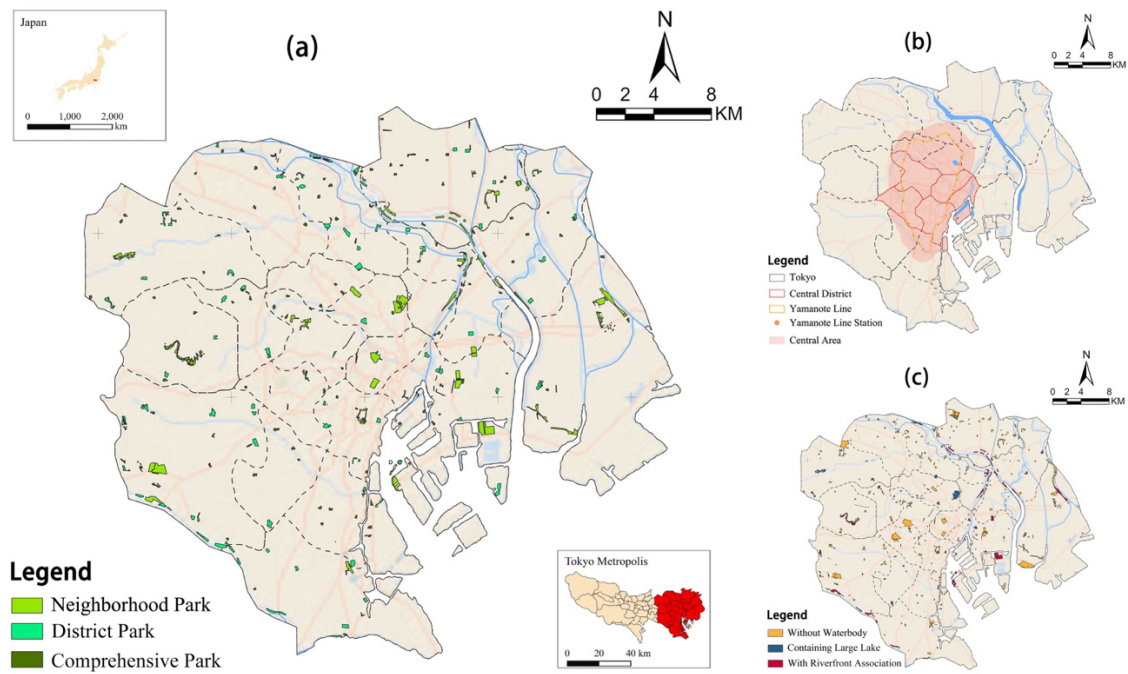

Figure 2. Study area and location of selected parks.

*Mobile Phone Data*

Mobile phone data for this study were provided by ZENRIN DataCom Co., LTD, utilizing the Konzatsu-Tokei® Data. The data captures aggregated people flows derived from individual location data sent from mobile phones under users' consent, through applications such as "docomomap navi" service (map navi, local guide, etc.) provided by NTT DoCoMo, INC, a major Japanese mobile phone operator (Guan et al., 2020). The original location dataset is GPS data sent in about every minimum period of 5 min and does not include the information to specify individual (Akiyama et al., 2013). Covering all areas of Japan, the dataset includes the entire year 2011 and around 9 billion mobile phone records. The spatial error is in general at a minimal level (within 15-30m without obstructions of high-rise buildings) for open space records. Although mobile phone data may be biased toward smartphone and application users (Ren et al., 2022), the rich spatiotemporal information it provides remains valuable for park rhythm research.

We employed a two-step process to identify park visitations. First, mobile phone GPS points were matched with park boundary polygons to detect spatial overlaps. This step was conducted by ZENRIN DataCom Co., LTD. Next, we excluded GPS points with dwell times more than 600 minutes or timestamps failing outside of standard park hours (6 a.m. to 10 p.m.), as these are likely to reflect home or work activities rather than park visits. Since our data provider defines a visit as a GPS point with a dwell time of more than five minutes (with shorter durations treated as pass-through traffic), we adopted the same five-minute threshold to identify park visits. A sensitivity analysis of this threshold is provided in Appendix B, Table S2. The final dataset includes 1,505,927 mobile phone records from 115,320 unique local users who visited at least one of the 254 selected parks in 2011. Table 1 displays a sample of the mobile phone data, which includes user ID, park ID, date of visit, a coordinates set (indicating stay points within the park), a time set (indicating timestamps of stay points in seconds regarding 0:00 a.m.), and travel distance from home to park (in kilometers). A summary of identified visitations by park categories and time periods is provided in Appendix B, Table S3.

Table 1. A sample of raw mobile phone data

| User ID | Park ID | Date | Coordinates set | Time set | Travel distance |
|---------|---------|------|-----------------|----------|-----------------|
| 482 | D065 | 2/8 | 139.722xxx : 35.631xxx \| 139.723xxx : 35.631xxx | 32439 \| 33893 | 2.45 |
| 1734 | C123 | 2/8 | 139.724xxx : 35.632xxx \| 139.724xxx : 35.632xxx | 48245 \| 54394 | 5.39 |
| 373 | D844 | 4/25 | 139.722xxx : 35.632xxx \| 139.722xxx : 35.632xxx \| 139.722xxx : 35.631xxx | 37483 \| 45839 \| 46724 | 12.48 |
| 3282 | B755 | 11/11 | 139.722xxx : 35.632xxx \| 139.722xxx : 35.632xxx | 47382 \| 50121 | 0.57 |

## *Analytical Methods for Temporal Classification*

### *Visitation dynamic curves (VDC)*

We propose a VDC-based approach to capture park visitation dynamics entailing multiple temporal dimensions. First, we divide each day into 30-minute intervals to assess fluctuations in visitor density, accounting for variations in park usage patterns throughout the operational hours (6 a.m. to 10 p.m.). Second, three typical days—a weekday, a weekend, and a holiday—are selected for analysis to reflect different visitation purposes and user groups. Finally, we examine these dynamics across four seasons—spring, summer, autumn, and winter—to understand seasonal variations. Typically, a park VDC consists of park visitor density across 96 time intervals, spanning three typical days within a specific season, which can be regarded as a time series (Brockwell & Davis, 1991) that directly reflects visitation dynamics and is easy to compare among parks and seasons. Further details on the data processing and examples of VDCs are expounded in Appendix C.

*Dynamic time warping (DTW) distance*

We employ dynamic time warping (DTW) distance (Müller, 2007) as a measure of similarity between park VDCs. Unlike Euclidean distance, which calculates the direct point-to-point distance and is often troubled by the "curse of dimensionality" (Verleysen & François, 2005), DTW distance allows for non-linear alignments between time series, making it more suitable for comparing activity patterns that vary in a continuous time scale (Chen et al., 2017). A detailed comparison between DTW and Euclidian distance is included in Appendix D.

We apply two clustering algorithms, K-means (Kanungo et al., 2002) and K-medoids (Park & Jun, 2009), chosen for their robust performance in unsupervised learning scenarios. We test different settings of classifications regarding distance measures (Euclidean or DTW), clustering algorithm (K-means or K-medoids), normalization methods (standardization or max-min scale, employed to eliminate redundant information before classification), and number of clusters (*K* ranging from two to ten). This multifaceted experiment ensures a comprehensive classification of urban parks discerning hyperparameter configurations. To assess the quality of clustering, we use the silhouette score (Nassif & Hruschka, 2013), a metric that evaluates how similar a point is to its own cluster compared to other clusters. The silhouette score ranges from -1 to 1, where a higher score indicates a clustering outcome that can better reflect the underlying patterns. By maximizing the silhouette score, we identify the classification that best represents the inherent patterns in park visitation dynamics. The calculation of silhouette score can be found in Appendix D.

*Park usage indicators, t-test, and disparity analysis*

We construct park usage indicators (PUIs) using mobile phone data to reflect visitor behaviors. Following previous research (Ren & Guan, 2022; Zhou et al., 2024), four primary dimensions are considered: visitation frequency, determined by averaging the number of unique user ID per park per day; visitor group, measured as the proportion of annual visitors (those who live farther than 20 km from a park and visit the park once a year) and regular visitors (those who live within 4km radius from the park and visit the park four times or more); travel distance, calculated as the average kilometers travelled to reach the park; and stay duration, representing the average time spent in the park measured in minutes. These indicators capture the features of actual park usage, aiding in testing the effectiveness of park classification.

T-test is employed as a statistical method to compare the PUIs between park categories. This test assesses whether visitor behaviors in one park category are statistically different from those in others, thereby checking the capability of park classification in reflecting the diversity of park usage. Moreover, we focus on the distribution equality of travel distance across different park categories, providing further insights of urban park access and resource allocation. The Gini coefficient, an equality measure that has been widely applied to evaluate income disparity (Gastwirth, 1972), is introduced as a quantifier of the inequality in park travel distances. The Lorenz curve, visualizing a distribution by plotting the cumulative percentage of mobile phone users against the cumulative percentage of travel distance, is employed alongside the Gini coefficient. Details of t-test, Gini coefficient, and Lorenz curve are included in Appendix E.

## Park Classification Results and Behavioral Insights

### *Six VDC-based Park Categories*

Figure 3 presents the the normalized VDCs for the six park clusters identified using K-means algorithm and DTW distance that result in the highest silhouette score. The silhouette scores across various clustering settings are reported in Appendix F.

Event-oriented parks (Cluster 1) account for 19.78% of the park VDCs and are characterized by a sharp increase in visitor density during holidays from 12 p.m. to 4 p.m. These parks maintain moderate levels of activity on weekends and lower usage on weekdays. The visitation pattern suggests that these parks serve as focal points for large-scale social or cultural events rather than daily use.

Weekend-centric parks (Cluster 2) constitute 20.67% of the park VDCs. These parks see a pronounced peak in visitor density on weekends between 10 a.m. and 4 p.m., catering to residents seeking leisure, recreation, or family outings on their days off. In contrast, weekday and holiday usage remains relatively low, suggesting these parks are not primary destinations for weekday activities or holiday events.

Commuter-accessible parks (Cluster 3) make up 10.43% of the park VDCs, featuring a distinctive weekday peak in visitor density between 10 a.m. and 2 p.m. This pattern suggests that these parks are frequented by office workers, commuters, and nearby residents who use them as convenient spaces for relaxation or exercise during work breaks. The weekday-centric use highlights their role in supporting daily routines and offering green spaces for brief, routine breaks amidst urban commutes.

Daily-leisure parks (Cluster 4) represent 11.02% of the park VDCs and exhibit a consistent, moderate level of visitor density throughout the week, with a slight midday peak.

These parks are popular for routine activities such as walking, jogging, and relaxing, appealing to local residents who incorporate them into their daily routines.

Balanced-usage parks (Cluster 5) account for 18.41% of the park VDCs, showcasing an evenly distributed pattern of visitor density across weekdays, weekends, and holidays. These parks maintain a consistent and relatively high level of activity throughout the week, with noticeable peaks in the early morning and late afternoon. They cater to a mix of both local residents and visitors engaging in varied activities and time preference, including exercise, leisure walks, and social gatherings.

Low-intensity parks (Cluster 6) contain 19.69% of the park VDCs, characterized by consistently low visitor density across all time periods. These parks are primarily frequented by a small number of local residents for brief walks or passive recreation, rather than serving as major destinations for leisure or social events. The subdued activity levels indicate that these parks are less central to urban life but to serve nearby communities more effectively.

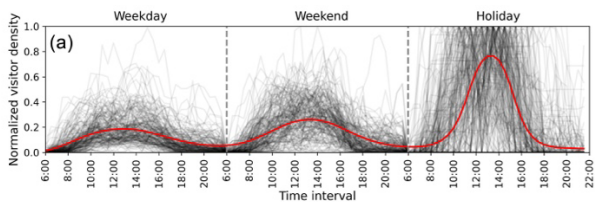

Cluster 1: Event-oriented parks (19.78%)

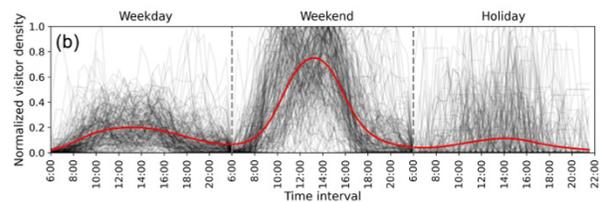

Cluster 2: Weekend-centric parks (20.67%)

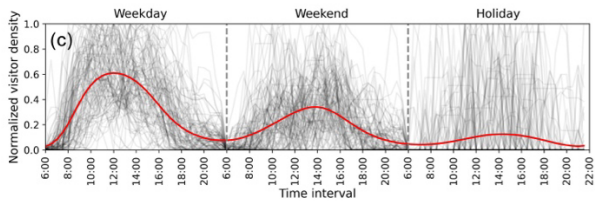

Cluster 3: Commuter-accessible parks (10.43%)

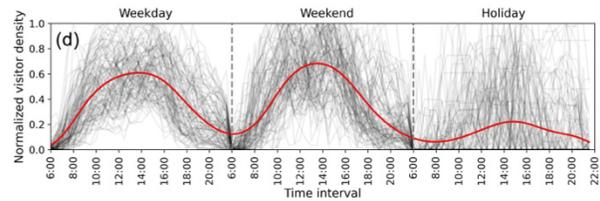

Cluster 4: Daily-leisure parks (11.02%)

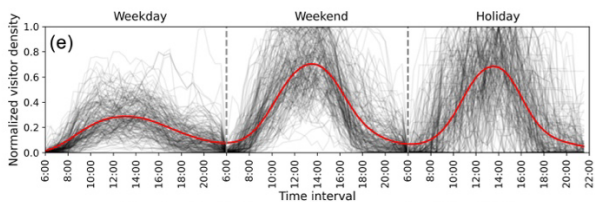

Cluster 5: Balanced-usage parks (18.41%)

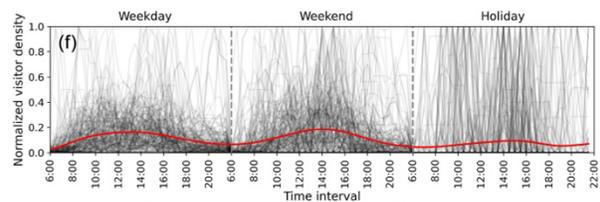

Cluster 6: Low-intensity parks (19.69%)

Figure 3. Classification results for K-Means and $K = 6$, whereby (a)-(f) depict the normalized visitor densities (y axis) by 30-minute time interval (x axis) for the six clusters.

Figure 4 presents the geographic distribution of the six park categories in four seasons. These categories are scattered throughout the city, which aligns with Tokyo's well-established and balanced urban park system. However, seasonal variation in park categories reveals an insightful dimension for understanding park usage. Event-oriented parks and balanced-usage parks are more prominent in spring (26.77% and 22.44%, respectively) and autumn (29.13% and 26.77%, respectively) than in summer (14.57% and 15.35%, respectively) and winter (9.84% and 7.87%, respectively). In contrast, low-intensity parks are more frequent in winter (27.95%) than spring (11.02%) and autumn (14.57%). Moreover, many parks displayed varied VDC categories across the four seasons, underscoring the importance of a seasonal perspective in park classification. Detailed proportions of these categories across seasons are presented in Appendix G.

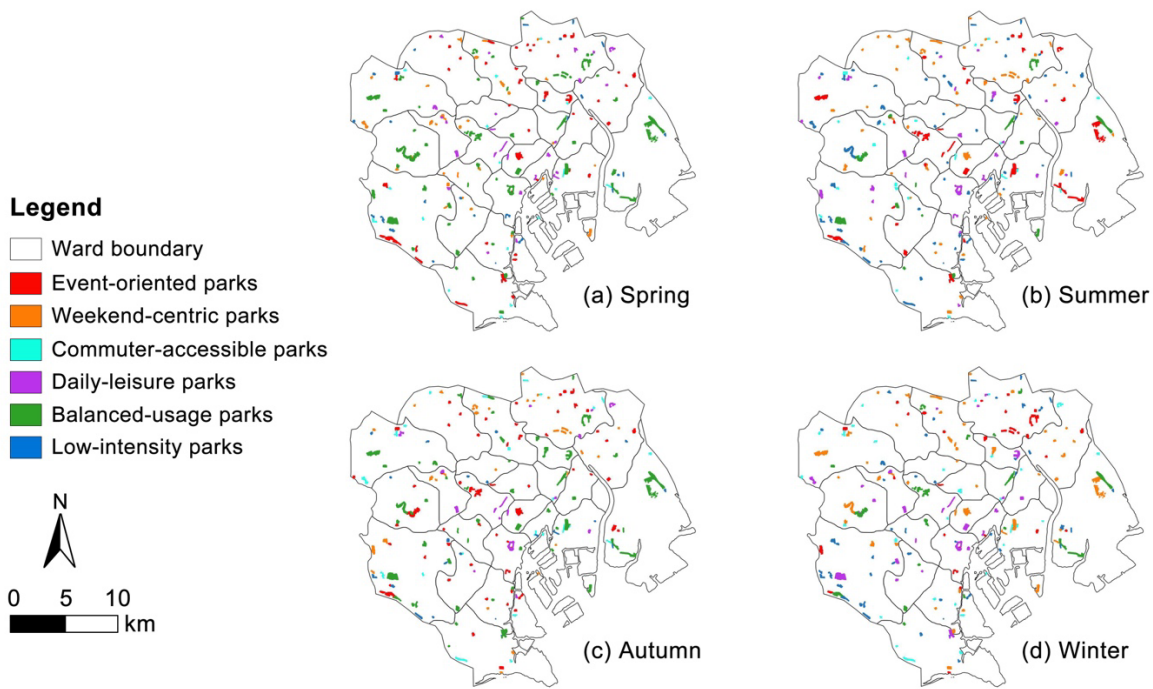

Figure 4. Geographic distribution of the six park categories.

***Insights from the Park Usage Perspective***

Figure 5 compares traditional park categories—based on size, location, and water feature integration—and the six novel categories derived from visitation dynamics. The results show that parks with similar physical characteristics can exhibit heterogenous visitation dynamics. For instance, within neighborhood parks, event-oriented parks account for 17.45%, weekend-centric parks for 21.76%, commuter-accessible parks for 15.11%, daily-leisure parks for 9.71%, balanced-usage parks account 8.63%, and low-intensity parks for 27.34%.

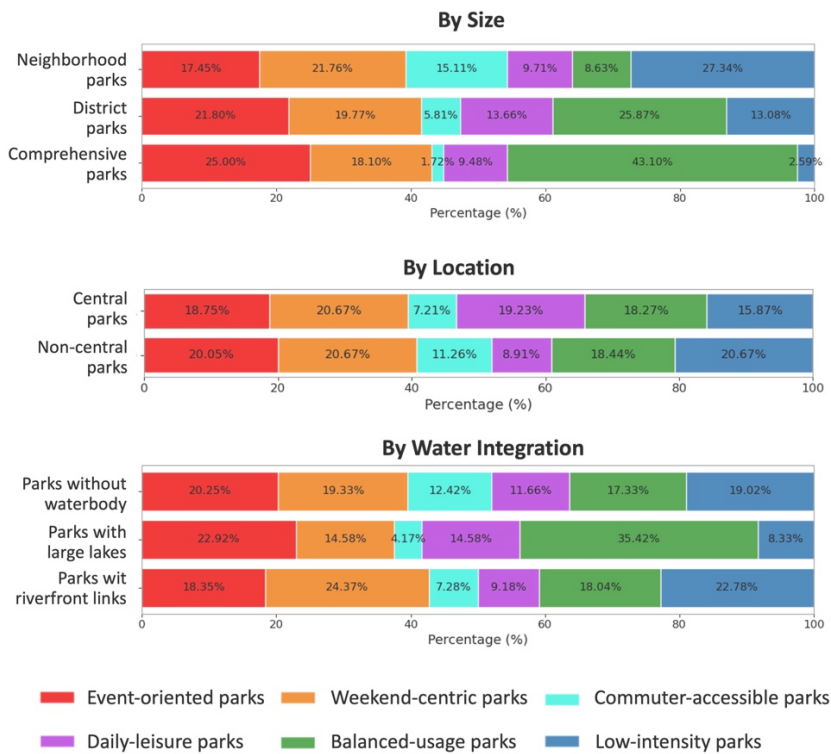

Figure 5. A comparison between traditional and VDC-based park categories.

Table 2 presents the results of t-test on PUIs across different park categories. The t-test analysis revealed significant variances in PUIs among the six VDC-based park categories. For instance, event-oriented parks show a markedly high visitation frequency (3,420 visitors per day), indicating their role as popular destinations for special events. In contrast, daily-leisure parks exhibit a high percentage of regular visitors (47.17%), suggesting that they cater primarily to nearby residents who visit frequently. Commuter-accessible parks demonstrate the longest average travel distance (11.06 km), reflecting their appeal to visitors traveling along commute routes from further afield. Low-intensity parks have the longest average stay duration (56.39 minutes), indicating that while they attract fewer visitors, the engagement level per visit is higher. These insights show that VDC-based classifications are more effective at capturing the diversity of park usage across multiple indicators, thereby providing more comprehensive behavioral insights for urban park planning and design.

Table 2. Results of t-test on PUIs per park category (each entry represents the mean of one indicator, and the number in the parenthesis is the standard error).

| Park category | Visitation frequency | % Annual visitors | % Regular visitors | Travel distance | Stay duration |
|---|---|---|---|---|---|
| All parks | 1,828 (150) | 13.49% (0.47%) | 42.10% (1.13%) | 9.38 (0.32) | 52.01 (1.21) |
| By park size | | | | | |
| Neighborhood parks | 853*** (53.23) | 13.45% (0.66%) | 45.97%*** (1.43%) | 8.63** (0.42) | 52.74 (1.78) |
| District parks | 2,128 (233.43) | 13.04% (0.77%) | 39.74% (2.02%) | 9.92 (0.57) | 51.26 (1.99) |
| Comprehensive parks | 5,612*** (735.64) | 15.07% (1.48%) | 30.58%*** (2.85%) | 11.43** (0.99) | 50.77 (2.54) |
| By park location | | | | | |
| Central parks | 1,786 (302.76) | 13.50% (1.15%) | 41.81% (2.50%) | 9.50 (0.70) | 52.23 (2.76) |
| Non-central parks | 1,839 (172.78) | 13.49% (0.52%) | 42.18% (1.27%) | 9.35 (0.37) | 51.96 (1.36) |
| By water integration | | | | | |
| Parks without waterbody | 1,590*** (120.47) | 12.91%** (0.40%) | 42.98% (0.96%) | 9.35 (0.29) | 52.18 (1.14) |
| Parks with large lakes | 1,828 (375.92) | 12.98% (1.28%) | 44.04% (3.39%) | 8.98 (0.96) | 54.16 (3.90) |
| Parks with riverfront links | 2,318*** (223.65) | 14.78%** (0.67%) | 39.99%* (1.53%) | 9.50 (0.41) | 51.35 (1.30) |
| By visitation dynamic curves (VDCs) | | | | | |
| Event-oriented parks | 3,420*** (327.86) | 16.05%*** (0.65%) | 36.15%*** (1.47%) | 10.65** (0.44) | 44.23*** (1.21) |
| Weekend-centric parks | 1,551 (76.85) | 13.45% (0.42%) | 39.98% (1.10%) | 8.86 (0.27) | 46.32*** (1.36) |
| Commuter-accessible parks | 1,626 (300.26) | 14.06% (0.73%) | 42.85% (1.80%) | 11.06*** (0.71) | 50.72 (1.92) |
| Daily-leisure parks | 1,407*** (129.52) | 13.69% (0.67%) | 47.17%*** (1.30%) | 9.56 (0.34) | 53.76* (1.44) |
| Balanced-usage parks | 2,683*** (239.39) | 13.88% (0.11%) | 38.06%*** (1.97%) | 10.07* (0.56) | 51.46 (1.30) |
| Low-usage parks | 1,286*** (85.93) | 10.91%*** (0.39%) | 48.29%*** (1.05%) | 7.95*** (0.30) | 56.39*** (1.22) |

Note: visitation frequency is in persons per park per day, travel distance is in kilometers, and stay duration is in minutes. '***' refers to p-value<0.01; '**' refers to p-value<0.05; '*' refers to p-value<0.1.

### *Insights from the Access Disparity Perspective*

The Gini coefficients and Lorenz curves in Figure 6 illustrate the disparity of travel distance variances across different park categories. We found that disparities in travel distances across VDC-based park categories are more pronounced than those classified by park size, location, or water integration. For instance, the Gini coefficient for VDC-based categories ranges from 0.469 (daily-leisure parks) to 0.543 (event-oriented parks), while categories based on physical characteristics exhibit lower Gini coefficients, such as 0.358 for comprehensive parks, 0.393 for central parks, and 0.434 for parks with riverfront links.

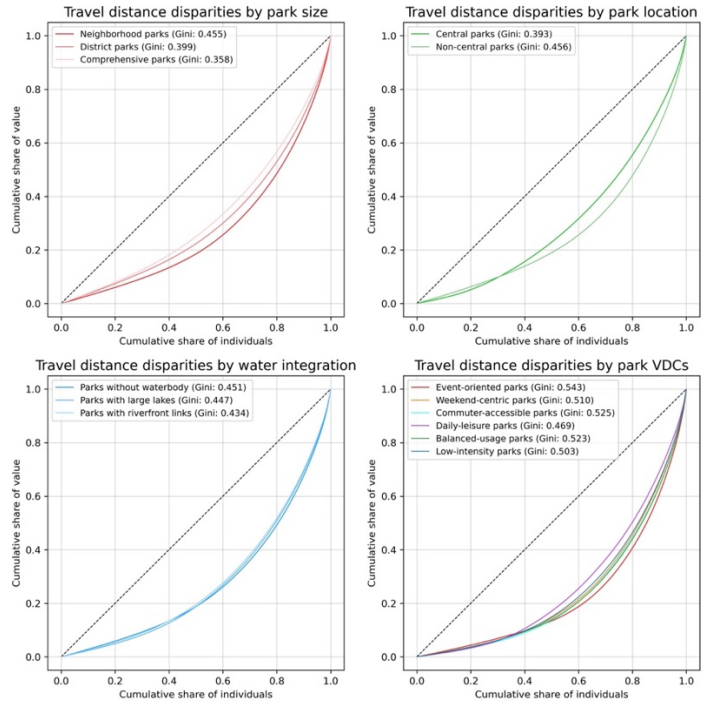

Figure 6. Travel distance disparities across park categories regarding physical factors and VDCs

Moreover, travel distance disparities across VDC-based park categories show significant seasonal variations (Figure 7). In spring and autumn, event-oriented parks display the highest Gini coefficients (0.694 in spring and 0.703 in autumn), followed by commuter-accessible parks (0.620 in spring and 0.578 in autumn) and weekend-centric parks (0.583 in spring and 0.655 in autumn). These values reflect substantial disparities in travel distance among visitors when the weather is suitable for outdoor activities, indicating that some visitors have to travel longer distances to access parks. In contrast, the disparities are less pronounced in summer and winter, with lower Gini coefficients, such as 0.494 for low-intensity parks in summer and 0.467 for daily-leisure parks in winter.

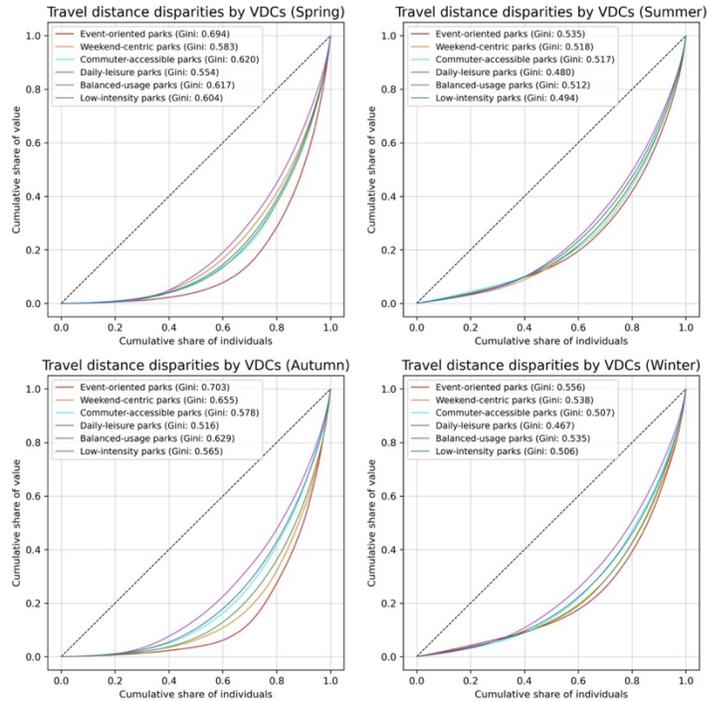

Figure 7. Travel distance disparities across VDC-based park categories in four seasons

Long travel distances observed in some park categories may reflect both intentional user choices and accessibility barriers (Bertram et al., 2017). To further distinguish access disparities that reflect inequities and those that reflect heterogeneity in use preferences, we introduce the concept of the "park detour ratio". This metric quantifies the additional distances visitors travelled compared to the distance to the nearest park in the same category from their home (see Appendix H). Table 3 reveals that park visitors do not always choose the nearest park (referring to variation in preferences), as indicated by park detour ratios generally greater than 1 across all VDC categories and seasons. Visitors also tend to travel farther during spring and autumn; for example, visitors travelled 2.18 times longer for balanced-usage parks in autumn, compared to 1.72 times in winter. Interestingly, the category with the highest disparity in distance (event-oriented parks) did not exhibit the highest detour ratios (lower than weekend-centric and balanced-usage parks), suggesting that accessibility barriers may still deter certain groups from reaching these parks. We need to mention that our analysis assumes that visitors depart from their home. Future research incorporating activity chain information would offer a more precise understanding of trip origins and motivations.

Table 3. Park detour ratio across VDC categories and seasons

| VDC-based park category | Spring | Summer | Autumn | Winter |
|---|---|---|---|---|
| Event-oriented parks | 1.87 | 1.74 | 1.92 | 1.64 |
| Weekend-centric parks | 1.99 | 2.01 | 2.06 | 1.89 |
| Commuter-accessible parks | 1.11 | 1.21 | 1.43 | 1.45 |
| Daily-leisure parks | 1.83 | 1.82 | 1.81 | 1.78 |
| Balanced-usage parks | 1.96 | 1.78 | 2.18 | 1.72 |
| Low-usage parks | 1.14 | 1.19 | 1.30 | 1.29 |

## Modes of Action for Temporal Park Planning

The findings from our study underscore the importance of integrating temporality and rhythmic into urban park planning. Planners, designers, and policymakers should seek to transform parks into resilient, inclusive spaces that respond to the evolving needs of diverse urban populations. This section outlines four key modes of action, providing a comprehensive framework for reimagining urban parks as rhythm-sensitive infrastructures that harmonize with the temporal complexities of urban life.

### *Tracking Park Rhythms Using Data-driven Techniques*

Tracking park rhythms forms the foundation of temporal park planning, as it provides empirical insight into how urban green spaces are used across time. Mobile phone data, with its high spatial-temporal resolution and broad geographic coverage, offers a scalable and cost-effective tool for monitoring park usage in real-time. Our case study in Tokyo demonstrates the feasibility of using such data to capture park rhythms at multiple temporal scales—daily, weekly, and seasonal. Distinct VDC patterns, such as midday peaks in commuter-accessible parks and holiday surges in event-oriented parks, offer a granular understanding of temporal dynamics that traditional place-based classifications overlook.

However, it is important to acknowledge the limitations of our mobile phone dataset. The reliance on active data collection (via app usage) introduces variability in sampling frequency and potential bias in representativeness (Chukwu et al., 2024). For example, visits by nearby residents or older individuals less likely to engage with location-based apps may be underrepresented. These limitations can explain the absence of early-morning peak in our identified visitation patterns. Future studies should integrate other geolocation data sources for cross validation. Environmental sensors can track microclimatic conditions (e.g., temperature, humidity) that influence visitation patterns (Cheung et al., 2021), while social media data can capture qualitative aspects of user experiences, such as emotional responses to park activities (Chuang et al., 2022). Additionally, satellite imagery and land use data offer a broader spatial context, revealing patterns of park usage in relation to urban development (Daunt & Silva, 2019). Cross-validating mobile phone data with these sources can mitigate its potential biases (Calabrese et al., 2013). The methodology proposed in this study—a standardized workflow for identifying VDC patterns—is not dependent on mobile phone data. Any dataset containing information on hourly park visits can be readily integrated into our framework. This adaptability allows researchers to apply the framework across diverse data sources and urban contexts, enabling robustness checks and broader comparative studies. By combining multiple data streams or using data in multiple years, planners can develop a more holistic and robust framework for tracking park rhythms.

The reliance on big data should not overshadow the value of traditional methods like surveys and field observations. Large-scale data often lacks contextual depth, such as the motivations behind park visits or the social interactions within these spaces. Surveys and ethnographic studies can provide such information, offering insights into user preferences, cultural practices, and unobserved barriers to access (Durán Vian et al., 2021; Li et al., 2021). Combining emerging big data with traditional thick data creates a synergistic approach, balancing scalability with nuanced understanding to ensure that temporal park planning is both data-driven and human-centered.

### Implementing Temporary Design for Dynamic Park Demands

The significant discrepancies in park usage indicators (PUIs) across VDC-based categories highlight the inadequacy of uniform park design. Instead, temporary design strategies that are adaptive to the unique temporal rhythms of each park category can improve functionality and user satisfaction. Below, we propose adaptive interventions for the six VDC-based park categories grounded in their distinct visitation patterns (Table 4).

*Event-Oriented Parks* require flexible infrastructure to manage holiday and event-driven visitation spikes. Modular installations such as temporary stages and vendor zones can accommodate festivals or markets (Romić & Šćitaroci, 2022), while dynamic space reconfiguration using retractable seating and movable barriers allows layouts to adapt to different event types (Carr & Dionisio, 2017). Tech-enabled crowd management tools, including real-time visitor tracking and capacity alerts, can mitigate overcrowding risks (Li et al., 2021). Collaboration with local authorities for traffic rerouting and emergency protocols can ensure safer, large-scale event execution.

*Balanced-Usage Parks*, with steady visitation across days, benefit from multi-functional zoning. Activity-specific areas—such as sports courts, quiet gardens, and cultural hubs—cater to diverse user groups, while seasonal programming (e.g., winter ice rinks or summer outdoor cinemas) sustains year-round engagement (Mensah et al., 2017). Connectivity enhancements, like pedestrian and cycling pathways integrated with adjacent neighborhoods, can improve accessibility and foster local interaction (Woo & Choi, 2022).

*Weekend-Centric Parks* demand family-centric amenities to address short but intensive weekend peaks. Temporary recreational hubs featuring BBQ pits, shaded picnic areas, and children's playgrounds cater to family outings (Bertram et al., 2017). Event-driven activations, such as guided nature walks or craft markets, can diversify offerings and manage crowd density (Loukaitou-Sideris & Sideris, 2009). Increased weekend stuffing can ensure cleanliness and safety without compromising the park's relaxed atmosphere.

*Commuter-Accessible Parks*, frequented by office workers, should prioritize convenience and efficiency. Quick-break infrastructure—modular seating, mobile charging stations, and micro-cafés—support short visits (Agustin et al., 2024), while shaded green nooks and mindfulness gardens offer stress relief during lunch breaks (Lin et al., 2013). Improved pedestrian links to nearby transit hubs help align park access with commuter rhythms, enhancing usability (Thakuriah et al., 2012)

*Daily-Leisure Parks*, serving local residents, thrive on routine-friendly features. Fitness-oriented designs—such as exercise equipment and walking trails—encourage daily physical activity (Duan et al., 2018), while community co-creation spaces (e.g., open-air pavilions) host neighborhood classes or gatherings (Peters et al., 2010). Seasonal landscaping rotations, including themed gardens or floral displays, sustain visual interest and visitation consistency (Kotsila et al., 2020).

*Low-Intensity Parks* should adopt dual strategies to balance preservation and activation. Enhancing tranquillity through native plantings, wildlife habitats, and secluded seating preserves their role as serene retreats (Brambilla et al., 2013). On the other hand, gentle activation via small-scale attractions—community art walls, vegetable plots, or storytelling boards—attracts visitors without overwhelming capacity (Watson et al., 2020).

Table 4. Temporal dynamics, visitor behavioral features, and adaptive strategies for six park categories.

| Park category | Temporal dynamics | Visitor behavioral features | Adaptive strategies |
|---|---|---|---|
| Event-Oriented Parks | Significant spikes in visitation during holidays | High frequency, more annual visitors, long travel distance, short stay duration | Install temporary seats; provide vendor zones; conduct traffic management; make emergency response plans |
| Balanced-Usage Parks | Consistent visitation throughout the week | High frequency, balanced visitor composition, long travel distance, moderate stay duration | Incorporate a variety of facilities to support different user groups; enhance connectivity with surrounding neighborhoods |
| Weekend-Centric Parks | Higher numbers of visitors on weekends | Moderate frequency, balanced visitor composition, moderate travel distance, short stay duration | Provide additional facilities that can accommodate family-oriented activities; increase staff presence during weekends |
| Commuter-Accessible Parks | Higher numbers of visitors on weekdays | Moderate frequency, balanced visitor composition, long travel distance, moderate stay duration | Provide facilities that support short, frequent visits; improve pedestrian and bicycle access; connect nearby public transport hubs |
| Daily-leisure parks | A steady visitor flow across weekdays and weekends | Low frequency, more regular visitors, moderate travel distance, moderate stay duration | Offer facilities such as fitness trails and areas for yoga or tai chi; provide covered areas for outdoor classes or community events |
| Low-intensity parks | Low usage during all time periods | Low frequency, more regular visitors, short travel distance, long stay duration | Focus on maintaining naturalistic elements; introduce small community-oriented facilities to attract more visitors |

*Establishing An Inclusive Park System for Spatiotemporal Equity*

Traditional park equity studies emphasize the geographic access equity, ensuring neighborhoods with diverse socio-demographics have similar proximity to green spaces (Xiao et al., 2019). However, our case study revealed that disparities in park accessibility may be underestimated when temporal dynamics are overlooked. By focusing solely on geographic distribution, planners risk perpetuating inequalities rooted in temporal constraints, such as available recreational time, access to mobility services, park operating hours, and weather conditions (Henckel & Thomaier, 2013). While spatial equity focuses on, "***Is there a park nearby?***" temporal equity directs attention to, "***Can people use a park when they need to?***"

We underscore the importance of integrating spatial and temporal equity in urban park planning. Figure 8 presents a framework illustrating how to bridge them together. From a top-down perspective, spatial park planning should center around place-based classifications, prioritizing physical attributes such as park size, location, and infrastructure. Planners should seek to ensure the equitable distribution of park resources across geographic areas and user groups through long-term budgeting and administrative regulation (Brown et al., 2014; Chen et al., 2018). From the bottom-up perspective, temporal park planning should leverage activity-based classifications to tailor park design to round-the-clock demands. For instance, parks may meet spatial equity criteria by being evenly located, but temporal equity requires that at least some of them remain accessible to late returnees at night, hanami participants on holidays, or outdoor enthusiasts in winter. This dual logic recognizes that parks should not only exist in space but also function in time.

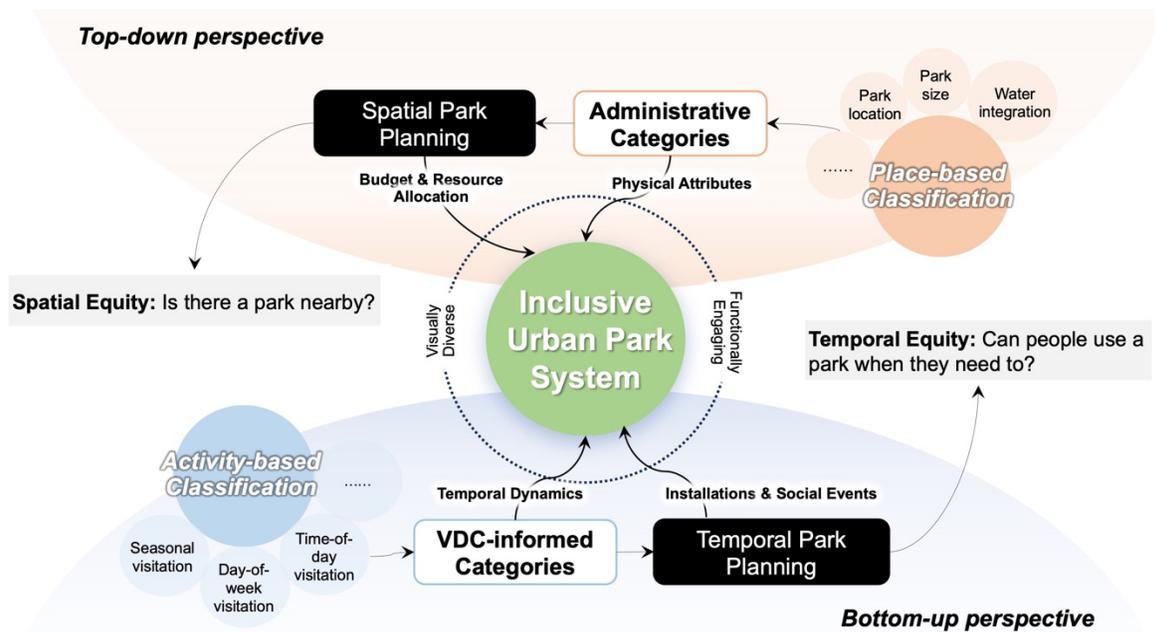

Figure 8. A framework of establishing an inclusive urban park system.

To operationalize this vision, we highlight three park categories that promote holistic park usage but were not identified in our case study[1]. First, neighborhood parks near residential areas can become ***Morning-Popular Parks*** by providing running tracks or fitness equipment for the elderly, and organizing morning activity programs (Sreetheran, 2017). Second, parks in central locations or near employment zones can extend their operating hours and enhance lighting to attract visitors at night, serving as ***Long-Evening Parks***. Third, green spaces sensitive to temperature can be transformed into ***Winter-Friendly Parks*** by offering sheltered seating areas, heated benches, and temporary indoor facilities to ensure human thermal comfort during snowy or cold weather (Sun & Sun, 2023). To check the effectiveness of these efforts, constructing innovative metrics to measure spatiotemporal equity could be a direction of future research.

---

[1] We did not find these park categories probably because our classification approach captured the significant fluctuations across typical days and seasons but overlooked subtle differences within the day.

### *Encouraging Community Engagement to Cultivate Time Identity*

Implementing temporal strategies in urban parks is inherently more complex than activity-based classification, as parks are profoundly shaped by their historical legacies, cultural traditions, and socio-demographic contexts (Durán Vian et al., 2021). Table S4 in Appendix I provides basic information and seasonal pictures of 12 selected parks from our case study, offering a more comprehensive view of how their temporal dynamics are formed. For instance, Shiba Park and Hibiya Park are comprehensive parks in central Tokyo, yet they belong to distinct activity-based categories across the seasons. Shiba Park is renowned for its cherry blossom viewings, along with its spacious grassy areas and seasonal flower displays, which are particularly attractive in the spring. In contrast, Hibiya Park is the venue for various outdoor events around the Open-Air Concert Hall and Pelican Fountain, which are frequently hosted in the summer but rare in the winter. These contextual nuances require temporal park planners to transform guiding principles into flexible, localized practice approaches.

Encouraging community engagement offers a pathway to address this complexity. By involving residents in the planning process, communities can develop park programming that aligns with local rhythms while fostering a shared sense of time identity. For example, a park in a neighborhood with a strong tradition of evening socializing might benefit from extended lighting and food stalls, while one in a family-oriented area could prioritize morning fitness programs and weekend markets. Such initiatives can make parks more responsive to temporal needs and strengthen the emotional connection between residents and their green spaces, embedding parks into the fabric of daily life.

Feasible strategies to encourage community engagement include participatory workshops, where residents co-design park activities and amenities (Shuib et al., 2015), and temporal stewardship programs, where volunteers take responsibility for seasonal maintenance or event organization (Daněk et al., 2020; Roman et al., 2015). Digital tools, such as mobile apps for real-time feedback or online platforms for collaborative planning, can further democratize participation (Kong et al., 2022; Torabi et al., 2020). However, these are only preliminary insights. Further studies are required to explore how different cultural, economic, and geographic contexts influence the effectiveness of such strategies. Onsite park case studies will provide more insights to make our parks not only temporally adaptive but also deeply rooted in the identities of the communities they serve.

**Conclusion**

In this study, we treated urban parks as dynamic, rhythmized spaces by integrating temporality into their classification and planning. Our findings offer several unique contributions to urban planning theory and practice. First, our temporal classification framework addresses a critical gap in park typologies by prioritizing when and how parks are used over static physical characteristics. This approach reveals behavioral nuances—such as midday commuter peaks, seasonal event surges, and persistent underuse in winter—that traditional models overlook. Second, integrating mobile phone data with spatial-temporal analysis advances methodological innovation, enabling scalable, real-time monitoring of park rhythms while highlighting disparities in travel distance and seasonal accessibility. Third, our proposed modes of action bridge theory and practice, advocating for adaptive strategies such as modular infrastructure for event-oriented parks, multi-functional zoning in balanced-usage parks, and community-driven activations in low-intensity parks. These strategies underscore the need to harmonize park design with the fluid temporal needs of diverse urban populations.

However, relying solely on mobile phone data presents unignorable limitations. These datasets may underrepresent certain populations—such as older adults, children, or individuals without smartphones—and often lack contextual detail about user motivations, social dynamics, or access barriers. This can result in biased interpretations of park usage and limit the inclusivity of data-driven planning. To address this, future research should integrate complementary data sources from urban sensing, surveys, and field observations to enhance data representativeness. In addition, challenges remain in scaling these planning strategies across cities with diverse cultural, climatic, and socio-economic contexts. Cross-city comparisons are needed to refine classification frameworks and test adaptive interventions in varied environments. Integrating emerging technologies such as AI-based crowd prediction, IoT-enabled microclimate monitoring, and participatory digital platforms can further enhance the precision and inclusivity of temporal planning.

Looking ahead, we envision a paradigm shift where urban parks evolve as living systems, continuously adapting to the rhythms of urban life. By embedding temporality into planning practice, planners, designers, and policymakers can transform parks into spaces that not only mitigate overcrowding and underuse but also cultivate a deeper sense of place and community. This shift will mark a new era in park planning—one where parks are not static amenities but dynamic infrastructures that harmonize with the pulse of the city, fostering sustainability, equity, and vitality for all.

# Reference

Adjei Mensah, C., Andres, L., Baidoo, P., Eshun, J. K., & Antwi, K. B. (2017). Community Participation in Urban Planning: The Case of Managing Green Spaces in Kumasi, Ghana. *Urban Forum*, *28*(2), 125–141. https://doi.org/10.1007/s12132-016-9295-7

Akiyama Y, Horanont T, & Shibasaki, R. (2013). Time-series analysis of visitors i n commercial areas using mass person trip data. *GIS Association of Japan: 22nd Annual Conference*.

Arti, R. S. & Jumadi. (2024). GIS Application For Mapping and Analyzing Urban Green Open Spaces (Case Study of Sragen City). *IOP Conference Series: Earth and Environmental Science*, *1357*(1), 012034. https://doi.org/10.1088/1755-1315/1357/1/012034

Bertram, C., Meyerhoff, J., Rehdanz, K., & Wüstemann, H. (2017). Differences in the recreational value of urban parks between weekdays and weekends: A discrete choice analysis. *Landscape and Urban Planning*, *159*, 5–14.

Brambilla, G., Gallo, V., & Zambon, G. (2013). The Soundscape Quality in Some Urban Parks in Milan, Italy. *International Journal of Environmental Research and Public Health*, *10*(6), 2348–2369. https://doi.org/10.3390/ijerph10062348

Bråten, L. N. (2024). Temporary temporariness? The (mis)use of tactical urbanism from the 'open city' framework. *Urban Studies*, 00420980241263436. https://doi.org/10.1177/00420980241263436

Brockwell, P. J., & Davis, R. A. (1991). *Time series: Theory and methods*. Springer science & business media.

Brown, G., Schebella, M. F., & Weber, D. (2014). Using participatory GIS to measure physical activity and urban park benefits. *Landscape and Urban Planning*, *121*, 34–44. https://doi.org/10.1016/j.landurbplan.2013.09.006

Calabrese, F., Diao, M., Di Lorenzo, G., Ferreira, J., & Ratti, C. (2013). Understanding

   individual mobility patterns from urban sensing data: A mobile phone trace example.

   *Transportation Research Part C: Emerging Technologies*, *26*, 301–313.

   https://doi.org/10.1016/j.trc.2012.09.009

Carr, J., & Dionisio, M. R. (2017). Flexible spaces as a "third way" forward for planning urban

   shared spaces. *Cities*, *70*, 73–82. https://doi.org/10.1016/j.cities.2017.06.009

Chen, X., & Hedayati Marzbali, M. (2024). How urban park features impact perceived safety by

   considering the role of time spent in the park, gender, and parental status. *Cities*, *153*,

   105272. https://doi.org/10.1016/j.cities.2024.105272

Chen, Y., Liu, X., Gao, W., Wang, R. Y., Li, Y., & Tu, W. (2018). Emerging social media data

   on measuring urban park use. *Urban Forestry & Urban Greening*, *31*, 130–141.

   https://doi.org/10.1016/j.ufug.2018.02.005

Chen, Y., Liu, X., Li, X., Liu, X., Yao, Y., Hu, G., Xu, X., & Pei, F. (2017). Delineating urban

   functional areas with building-level social media data: A dynamic time warping (DTW)

   distance based k -medoids method. *Landscape and Urban Planning*, *160*, 48–60.

   https://doi.org/10.1016/j.landurbplan.2016.12.001

Cheung, P. K., Jim, C. Y., & Siu, C. T. (2021). Effects of urban park design features on summer

   air temperature and humidity in compact-city milieu. *Applied Geography*, *129*, 102439.

   https://doi.org/10.1016/j.apgeog.2021.102439

Chuang, I.-T., Benita, F., & Tunçer, B. (2022). Effects of urban park spatial characteristics on

   visitor density and diversity: A geolocated social media approach. *Landscape and Urban

   Planning*, *226*, 104514. https://doi.org/10.1016/j.landurbplan.2022.104514

Chukwu, M., Huang, X., Wang, S., Li, X., & Wei, H. (2024). Urban park accessibility

    assessment using human mobility data: A systematic review. *Annals of GIS*, *30*(2), 181–

    198. https://doi.org/10.1080/19475683.2024.2341700

Daněk, J., Zelený, J., Pecka Sejková, A., & Vačkářů, D. (2020). Exploring and Visualizing

    Stakeholder Value Regimes in the Context of Peri-Urban Park Planning. *Society &*

    *Natural Resources*, *33*(7), 927–940. https://doi.org/10.1080/08941920.2019.1688440

Duan, Y., Wagner, P., Zhang, R., Wulff, H., & Brehm, W. (2018). Physical activity areas in

    urban parks and their use by the elderly from two cities in China and Germany.

    *Landscape and Urban Planning*, *178*, 261–269.

    https://doi.org/10.1016/j.landurbplan.2018.06.009

Durán Vian, F., Pons Izquierdo, J. J., & Serrano Martínez, M. (2021). River-city recreational

    interaction: A classification of urban riverfront parks and walks. *Urban Forestry &*

    *Urban Greening*, *59*, 127042. https://doi.org/10.1016/j.ufug.2021.127042

Dynamic Time Warping. (2007). In Meinard Müller, *Information Retrieval for Music and*

    *Motion* (pp. 69–84). Springer Berlin Heidelberg. https://doi.org/10.1007/978-3-540-

    74048-3_4

Erdmann-Goldoni, C. (2024). Contemporary Art in Public Spaces: Forms of Expression, Social

    Significance, and Revitalization. *European Public & Social Innovation Review*, *9*, 1–20.

    https://doi.org/10.31637/epsir-2024-867

Ferrer-Ortiz, C., Marquet, O., Mojica, L., & Vich, G. (2022). Barcelona under the 15-Minute

    City Lens: Mapping the Accessibility and Proximity Potential Based on Pedestrian Travel

    Times. *Smart Cities*, *5*(1), 146–161. https://doi.org/10.3390/smartcities5010010

Gastwirth, J. L. (1972). The Estimation of the Lorenz Curve and Gini Index. *The Review of Economics and Statistics*, *54*(3), 306. https://doi.org/10.2307/1937992

Grilli, G., Mohan, G., & Curtis, J. (2020). Public park attributes, park visits, and associated health status. *Landscape and Urban Planning*, *199*. https://doi.org/10.1016/j.landurbplan.2020.103814

Guan, C. H., Song, J., Keith, M., Akiyama, Y., Shibasaki, R., & Sato, T. (2020). Delineating urban park catchment areas using mobile phone data: A case study of Tokyo. *Computers, Environment and Urban Systems*, *81*, 101474.

Guan, C. H., Song, J., Keith, M., Zhang, B., Akiyama, Y., Da, L., Shibasaki, R., & Sato, T. (2021). Seasonal variations of park visitor volume and park service area in Tokyo: A mixed-method approach combining big data and field observations. *Urban Forestry & Urban Greening*, *58*, 126973.

Guan, C., & Zhou, Y. (2024). Exploring environmental equity and visitation disparities in peri-urban parks: A mobile phone data-driven analysis in Tokyo. *Landscape and Urban Planning*, *248*, 105104. https://doi.org/10.1016/j.landurbplan.2024.105104

Hägerstrand, T. (1970). What about people in Regional Science? *Papers of the Regional Science Association*, *24*(1), 6–21. https://doi.org/10.1007/BF01936872

Henckel, D., & Thomaier, S. (2013). Efficiency, Temporal Justice, and the Rhythm of Cities. In D. Henckel, S. Thomaier, B. Könecke, R. Zedda, & S. Stabilini (Eds.), *Space–Time Design of the Public City* (pp. 99–117). Springer Netherlands. https://doi.org/10.1007/978-94-007-6425-5_8

Ibes, D. C. (2015). A multi-dimensional classification and equity analysis of an urban park

system: A novel methodology and case study application. *Landscape and Urban

Planning*, *137*, 122–137.

Imma Widyawati Agustin, Ismu Rini Dwi Ari, Irawati, S., & Siankwilimba, E. (2024). Strategy

for Implementing Park-and-Ride as a Supporting Facility for Commuter Movement.

*Evergreen*, *11*(2), 1068–1080. https://doi.org/10.5109/7183406

Jabbar, M., Yusoff, M. M., & Shafie, A. (2022). Assessing the role of urban green spaces for human well-being: A systematic review. *GeoJournal*, *87*(5), 4405–4423.

https://doi.org/10.1007/s10708-021-10474-7

Kanungo, T., Mount, D. M., Netanyahu, N. S., Piatko, C. D., Silverman, R., & Wu, A. Y. (2002).

An efficient k-means clustering algorithm: Analysis and implementation. *IEEE

Transactions on Pattern Analysis and Machine Intelligence*, *24*(7), 881–892.

https://doi.org/10.1109/TPAMI.2002.1017616

Ke, L., Furuya, K., & Luo, S. (2021). Case comparison of typical transit-oriented-development stations in Tokyo district in the context of sustainability: Spatial visualization analysis based on FAHP and GIS. *Sustainable Cities and Society*, *68*.

https://doi.org/10.1016/j.scs.2021.102788

Kong, L., Liu, Z., Pan, X., Wang, Y., Guo, X., & Wu, J. (2022). How do different types and

landscape attributes of urban parks affect visitors' positive emotions? *Landscape and

Urban Planning*, *226*, 104482. https://doi.org/10.1016/j.landurbplan.2022.104482

Kotsila, P., Hörschelmann, K., Anguelovski, I., Sekulova, F., & Lazova, Y. (2020). Clashing

temporalities of care and support as key determinants of transformatory and justice potentials in urban gardens. *Cities*, *106*, 102865.

https://doi.org/10.1016/j.cities.2020.102865

Krmpotić Romić, I., & Bojanić Obad Šćitaroci, B. (2022). Temporary Urban Interventions in

Public Space. *Prostor*, *30*(2(64)), 178–187. https://doi.org/10.31522/p.30.2(64).4

Lefebvre, H. (2004). *Rhythmanalysis: Space, time and everyday life*. Continuum.

Lefebvre, H. (2013). *Rhythmanalysis: Space, time and everyday life*. Bloomsbury Publishing.

Li, F., Yao, N., Liu, D., Liu, W., Sun, Y., Cheng, W., Li, X., Wang, X., & Zhao, Y. (2021).

Explore the recreational service of large urban parks and its influential factors in city

clusters – Experiments from 11 cities in the Beijing-Tianjin-Hebei region. *Journal of*

*Cleaner Production*, *314*, 128261. https://doi.org/10.1016/j.jclepro.2021.128261

Liang, H., & Zhang, Q. (2021). Temporal and spatial assessment of urban park visits from

multiple social media data sets: A case study of Shanghai, China. *Journal of Cleaner*

*Production*, *297*, 126682.

Lin, T.-P., Tsai, K.-T., Liao, C.-C., & Huang, Y.-C. (2013). Effects of thermal comfort and adaptation on park attendance regarding different shading levels and activity types.

*Building and Environment*, *59*, 599–611. https://doi.org/10.1016/j.buildenv.2012.10.005

Liu, K., Murayama, Y., & Ichinose, T. (2021). A multi-view of the daily urban rhythms of human mobility in the Tokyo metropolitan area. *Journal of Transport Geography*, *91*,

102985. https://doi.org/10.1016/j.jtrangeo.2021.102985

Loukaitou-Sideris, A., Levy-Storms, L., Chen, L., & Brozen, M. (2016). Parks for an Aging

Population: Needs and Preferences of Low-Income Seniors in Los Angeles. *Journal of*

*the American Planning Association*, *82*(3), 236–251.

https://doi.org/10.1080/01944363.2016.1163238

Loukaitou-Sideris, A., & Sideris, A. (2009). What Brings Children to the Park? Analysis and

Measurement of the Variables Affecting Children's Use of Parks. *Journal of the*

*American Planning Association*, *76*(1), 89–107.

https://doi.org/10.1080/01944360903418338

Lynch, K. (1976). *What time is this place?* MIT press.

Martin, M., Deas, I., & Hincks, S. (2019). The Role of Temporary Use in Urban Regeneration:

Ordinary and Extraordinary Approaches in Bristol and Liverpool. *Planning Practice &*

*Research*, *34*(5), 537–557. https://doi.org/10.1080/02697459.2019.1679429

Monz, C., Mitrovich, M., D'Antonio, A., & Sisneros-Kidd, A. (2019). Using Mobile Device

Data to Estimate Visitation in Parks and Protected Areas: An Example from the Nature

Reserve of Orange County, California. *Journal of Park and Recreation Administration*.

https://doi.org/10.18666/JPRA-2019-9899

Mulíček, O., Osman, R., & Seidenglanz, D. (2015). Urban rhythms: A chronotopic approach to urban timespace. *Time & Society*, *24*(3), 304–325.

https://doi.org/10.1177/0961463X14535905

Nassif, L. F. D. C., & Hruschka, E. R. (2013). Document Clustering for Forensic Analysis: An

Approach for Improving Computer Inspection. *IEEE Transactions on Information*

*Forensics and Security*, *8*(1), 46–54. https://doi.org/10.1109/TIFS.2012.2223679

Osman, R., & Mulíček, O. (2017). Urban chronopolis: Ensemble of rhythmized dislocated

places. *Geoforum*, *85*, 46–57. https://doi.org/10.1016/j.geoforum.2017.07.013

Park, H.-S., & Jun, C.-H. (2009). A simple and fast algorithm for K-medoids clustering. *Expert*

*Systems with Applications*, *36*(2), 3336–3341. https://doi.org/10.1016/j.eswa.2008.01.039

Peters, K., Elands, B., & Buijs, A. (2010). Social interactions in urban parks: Stimulating social

cohesion? *Urban Forestry & Urban Greening*, *9*(2), 93–100.

https://doi.org/10.1016/j.ufug.2009.11.003

Petryshyn, N. (2022, February 2). A City for All Seasons: Winter City Planning for Toronto's

Parks and Public Spaces. *Ryerson University*. https://doi.org/10.32920/19083320.v1

Pfeiffer, D., Ehlenz, M. M., Andrade, R., Cloutier, S., & Larson, K. L. (2020). Do Neighborhood

Walkability, Transit, and Parks Relate to Residents' Life Satisfaction?: Insights From

Phoenix. *Journal of the American Planning Association*, *86*(2), 171–187.

https://doi.org/10.1080/01944363.2020.1715824

Pierri Daunt, A. B., & Sanna Freire Silva, T. (2019). Beyond the park and city dichotomy: Land

use and land cover change in the northern coast of São Paulo (Brazil). *Landscape and

Urban Planning*, *189*, 352–361. https://doi.org/10.1016/j.landurbplan.2019.05.003

Ren, X., & Guan, C. (2022). Evaluating geographic and social inequity of urban parks in

Shanghai through mobile phone-derived human activities. *Urban Forestry & Urban

Greening*, *76*, 127709.

Ren, X., Guan, C., Wang, D., Yang, J., Zhang, B., & Keith, M. (2022). Exploring land use

functional variance using mobile phone derived human activity data in Shanghai.

*Environment and Planning B: Urban Analytics and City Science*, 23998083221103261.

Rigolon, A., & Németh, J. (2021). What Shapes Uneven Access to Urban Amenities? Thick

Injustice and the Legacy of Racial Discrimination in Denver's Parks. *Journal of Planning

Education and Research*, *41*(3), 312–325. https://doi.org/10.1177/0739456X18789251

Roman, L. A., Walker, L. A., Martineau, C. M., Muffly, D. J., MacQueen, S. A., & Harris, W.

(2015). Stewardship matters: Case studies in establishment success of urban trees. *Urban

*Forestry & Urban Greening*, *14*(4), 1174–1182.

https://doi.org/10.1016/j.ufug.2015.11.001

Setagaya Ward Government. (2016). *Setagaya Ward Open Space Record (In Japanese: 世田谷 区都市公園等調査)*. http://www.city.setagaya.lg.jp/kurashi/ 102/126/419/410/d00019129.html.

Shuib, K. B., Hashim, H., & Nasir, N. A. M. (2015). Community Participation Strategies in Planning for Urban Parks. *Procedia - Social and Behavioral Sciences*, *168*, 311–320. https://doi.org/10.1016/j.sbspro.2014.10.236

Song, Y., Zhang, L., Dong, X., & Zhang, M. (2024). Can environmental regulation foster incremental enhancement and quality improvement in green technological innovation under the background of the digital development? *Environment, Development and Sustainability*. https://doi.org/10.1007/s10668-024-04862-5

Sreetheran, M. (2017). Exploring the urban park use, preference and behaviours among the residents of Kuala Lumpur, Malaysia. *Urban Forestry & Urban Greening*, *25*, 85–93. https://doi.org/10.1016/j.ufug.2017.05.003

Stevens, Q., Leorke, D., Dovey, K., Awepuga, F., & Morley, M. (2024). From 'pop-up' to permanent: Temporary urbanism as an emerging mode of strategic open-space planning. *Cities*, *154*, 105376. https://doi.org/10.1016/j.cities.2024.105376

Sun, P., & Sun, W. (2023). Retracted: Data-Driven Winter Landscape Design and Pleasant Factor Analysis of Elderly Friendly Parks in Severe Cold Cities in Northeast China under the Background of Artificial Intelligence. *Security and Communication Networks*, *2023*, 1–1. https://doi.org/10.1155/2023/9821850

Thakuriah, P. (Vonu), Metaxatos, P., Lin, J., & Jensen, E. (2012). An examination of factors affecting propensities to use bicycle and pedestrian facilities in suburban locations. *Transportation Research Part D: Transport and Environment*, *17*(4), 341–348. https://doi.org/10.1016/j.trd.2012.01.006

Tokyo Metropolitan Government. (2023). *Tokyo's History, Geography, and Population*. https://www.metro.tokyo.lg.jp/english/about/history/history02.html

Torabi, N., Lindsay, J., Smith, J., Khor, L.-A., & Sainsbury, O. (2020). Widening the lens: Understanding urban parks as a network. *Cities*, *98*, 102527. https://doi.org/10.1016/j.cities.2019.102527

Veitch, J., Ball, K., Rivera, E., Loh, V., Deforche, B., & Timperio, A. (2021). Understanding children's preference for park features that encourage physical activity: An adaptive choice based conjoint analysis. *International Journal of Behavioral Nutrition and Physical Activity*, *18*(1), 133. https://doi.org/10.1186/s12966-021-01203-x

Verleysen, M., & François, D. (2005). The Curse of Dimensionality in Data Mining and Time Series Prediction. In J. Cabestany, A. Prieto, & F. Sandoval (Eds.), *Computational Intelligence and Bioinspired Systems* (Vol. 3512, pp. 758–770). Springer Berlin Heidelberg. https://doi.org/10.1007/11494669_93

Walter, M., Bagozzi, B. E., Ajibade, I., & Mondal, P. (2023). Social media analysis reveals environmental injustices in Philadelphia urban parks. *Scientific Reports*, *13*(1), 12571. https://doi.org/10.1038/s41598-023-39579-4

Wang, P., Zhou, B., Han, L., & Mei, R. (2021). The motivation and factors influencing visits to small urban parks in Shanghai, China. *Urban Forestry & Urban Greening*, *60*, 127086. https://doi.org/10.1016/j.ufug.2021.127086

Watson, C. J., Carignan-Guillemette, L., Turcotte, C., Maire, V., & Proulx, R. (2020). Ecological and economic benefits of low-intensity urban lawn management. *Journal of Applied Ecology*, *57*(2), 436–446. https://doi.org/10.1111/1365-2664.13542

Wolch, J. R., Byrne, J., & Newell, J. P. (2014). Urban green space, public health, and environmental justice: The challenge of making cities 'just green enough.' *Landscape and Urban Planning*, *125*, 234–244.

Woo, J. H., & Choi, H. (2022). A trade-off method through connectivity analysis applied for sustainable design and planning of large urban parks. *International Journal of Sustainable Development & World Ecology*, *29*(2), 139–152. https://doi.org/10.1080/13504509.2021.1931982

Wunderlich, F. M. (2013). Place-Temporality and Urban Place-Rhythms in Urban Analysis and Design: An Aesthetic Akin to Music. *Journal of Urban Design*, *18*(3), 383–408. https://doi.org/10.1080/13574809.2013.772882

Wunderlich, F. M. (2023). *Temporal Urban Design: Temporality, Rhythm and Place*. Taylor & Francis.

Xiao, Y., Wang, D., & Fang, J. (2019). Exploring the disparities in park access through mobile phone data: Evidence from Shanghai, China. *Landscape and Urban Planning*, *181*, 80–91.

Zhai, Y., Li, D., Wu, C., & Wu, H. (2021). Urban park facility use and intensity of seniors' physical activity – An examination combining accelerometer and GPS tracking. *Landscape and Urban Planning*, *205*, 103950.

Zhang, R., Sun, F., Shen, Y., Peng, S., & Che, Y. (2021). Accessibility of urban park benefits with different spatial coverage: Spatial and social inequity. *Applied Geography*, *135*, 102555. https://doi.org/10.1016/j.apgeog.2021.102555

Zhang, R., Zhang, C.-Q., Cheng, W., Lai, P. C., & Schüz, B. (2021). The neighborhood socioeconomic inequalities in urban parks in a High-density City: An environmental justice perspective. *Landscape and Urban Planning*, *211*, 104099. https://doi.org/10.1016/j.landurbplan.2021.104099

Zhou, K., Song, Y., & Tan, R. (2021). Public perception matters: Estimating homebuyers' willingness to pay for urban park quality. *Urban Forestry & Urban Greening*, *64*, 127275. https://doi.org/10.1016/j.ufug.2021.127275

Zhou, Y., Guan, C., Wu, L., Li, Y., Nie, X., Song, J., Kim, S. K., & Akiyama, Y. (2024). Visitation-based classification of urban parks through mobile phone big data in Tokyo. *Applied Geography*, *167*, 103300. https://doi.org/10.1016/j.apgeog.2024.103300